\documentclass[letterpaper,11pt,leqno]{article}
\usepackage{paper}
\hypersetup{pdftitle={Fake it till you make it: using artificial turbulence to achieve swift converged turbulence statistics in a pressure-driven channel flow}}

\begin{document}

\title{Fake it till you make it: using artificial turbulence to achieve swift converged turbulence statistics in a pressure-driven channel flow}

\author{Akshay Patil{{\href{https://orcid.org/0000-0001-9807-0733}{\includegraphics{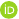}}}} and Clara Garc\'{i}a-S\'{a}nchez{{\href{https://orcid.org/0000-0002-5355-4272}{\includegraphics{orcid_logo}}}}}

\date{Nov. 2024}   


\begin{titlepage}
\maketitle

In this study, we introduced a simple yet innovative method to trigger turbulence in a channel flow to achieve statistically stationary flow conditions. 
We compare this new method based on synthetically generated three-dimensional turbulence with two other well-established methods, namely, linear profile superposed with random noise and descending counter-rotating vortices and log-law profile superposed with random noise and descending counter-rotating vortices. We found that synthetically generated three-dimensional turbulence provides a computationally cheap and effective way to reduce simulation spin-up to achieve statistically stationary flow conditions when a precursor turbulent initial condition is not available. 
At a one-time cost of less than 1~CPU hour to generate the synthetic turbulent initial condition, the flow becomes statistically stationary within 3 eddy turnovers for all the parameters of interest in wall-bounded pressure-driven channel flow simulations when compared to other alternatives that can take more than 10 eddy turnovers resulting in substantial savings in the computational cost.

\end{titlepage}

\section{Introduction}\label{s:introduction}
 
Pressure-driven turbulent channel flows have significantly improved our understanding of wall-bounded turbulence as detailed by \citep{JimenezMoin1991,KasagiTomitaKuroda1992,MoserKimMansour1999,Marusicetal2010,Jimenez2011,Smits2013,LozanoDuranetal2020,Jieetal2022}, to list a few. While such studies provide a fundamentally novel insight into flow physics, the computational resources required to simulate such flows can become increasingly demanding as the flow Reynolds number increases. Recent estimates by \cite{Horwitz2024} suggest that a turbulent channel flow simulated by \cite{VelaMatrinetal2021}\footnote{This case is only listed as an example for illustration purposes.} using 512 Graphics Processing Units (GPU) can use up to $5.98 \times 10^5$ kilo-Watt-hours (kWh) of energy and emit an equivalent of $6894$ kg of CO$_2$. The simulation by \cite{VelaMatrinetal2021} solved for the three-dimensional flow field at a friction Reynolds number ($Re_{\tau} \equiv u_{\tau} H/\nu$) of 5303, where $u_{\tau}$ is the friction velocity, $H$ is the channel half-height, and $\nu$ is the kinematic viscosity of the fluid. The total wall-clock time for this simulation was $1.4 \times 10^6$ CPU-hours, resulting in a total emission of $\sim 2.52$~kg of CO$_2$ per wall clock-hour \citep{Horwitz2024}. As illustrated by this simple estimate, this can pose a significant hurdle when statistically stationary flow conditions are to be achieved that require long simulation spin-up times before flow parameters can be averaged.

Without appropriate initial conditions to begin the channel flow simulations, most simulation frameworks generate three-dimensional flow fields initialised using simple analytical profiles for the streamwise velocity components superposed with white noise \cite{NelsonFringer2017}. In some studies, a pair of counter-rotating vortices \cite{HenningsonKim1991} is added to the linear-log-law profile to trigger the transition to turbulence and accelerate momentum mixing in the vertical direction, thus reducing the CPU time spent arriving at statistically stationary flow conditions \citep{Costa2018}. However, as discussed by \cite{NelsonFringer2017} and \cite{Costa2018}, the simulations require approximately 10 turnover ($T_{\epsilon} \equiv H/u_{\tau}$) periods to reach stationary flow conditions. This computational cost can scale to a drastically large amount with increasing flow Reynolds number, which can be otherwise spent on collecting valuable statistics to support the requisite inferences. In this study, we used synthetically generated three-dimensional turbulent flow field as initial conditions for two $Re_{\tau}$ in a pressure-driven channel flow to understand how the spin-up time compares with existing methods to achieve stationary flow conditions. The focus is to understand the effect of time-marching a given initial condition without any additional forcing apart from the constant pressure gradient that drives the flow. The following sections discuss the governing equations and numerical methods used to solve the flow equations and generate the initial conditions. This is followed by a detailed discussion of the results obtained from the various methods and comparing the quality of the statistics obtained.

\section{Governing Equations and Numerical Methods}

In this study, we consider the non-dimensional incompressible Navier-Stokes momentum equations given by

\begin{equation} \label{eq:NSeq}
    \partial_t^* u_i^* + \partial_j^* u_j^* u_i^* = - \partial_i^* p^* + {Re_{\tau}}^{-1} \partial_j^* \partial_j^* u_i^* + \Pi_c \delta_{i1}, 
\end{equation}

\noindent subject to the incompressible continuity equation given by

\begin{equation} \label{eq:continuity}
    \partial_i^* u_i^* = 0,
\end{equation}

\noindent where $t$ is time, $u_i$ is the velocity vector, $p$ is pressure, $Re_{\tau} \equiv u_{\tau} H / \nu$ is the Reynolds number, $u_{\tau}$ is the friction velocity, $H$ is the channel height, $\nu$ is the kinematic viscosity of the fluid, $\Pi_c$ is the driving pressure gradient, and $\delta_{ij}$ is the Kronecker delta function. The $( \cdot )^*$ notation represents the non-dimensional parameters obtained using $u_{\tau}$ and $H$.  Using this choice of non-dimensionalisation, the driving pressure gradient is exactly unity.
The governing equations are numerically integrated using the open-source massively parallel {CaNS} solver developed by \cite{Costa2018}. 
{CaNS} solves the governing equations using a second-order accurate spatial discretisation and a third-order accurate temporal discretisation using the low-storage Runge-Kutta 3-step method using the fractional step algorithm \citep{KimMoin1985}. 
The flow variables are arranged on a staggered grid where scalars are placed at the cell centre while vector components are located at the cell faces \citep{FerzigerPericStreet2019}.
CaNS has been extensively validated for channel flow simulations, and further details can be found in \cite{Costa2018} and will not be discussed for the sake of brevity. 

For all the cases discussed in this work, the channel has dimensions $L_{x_1} \times L_{x_2} \times L_{x_3} \equiv 4\pi H \times 2\pi H \times H$, where $x_i$ corresponds to the coordinate axes in streamwise, spanwise, and vertical directions, respectively. 
The flow is driven by a constant pressure gradient $\Pi_c = 1.0$ subject to periodic boundary conditions in the streamwise and spanwise directions, and no-slip boundary condition at $x_3 = 0$ and free slip boundary where $\partial_3 u_1 = \partial_3 u_2 = 0$ and $u_3 = 0$ at $x_3 = H$. 
The flow field was initialised using three different initial conditions: namely, inverse linear profile superposed with white noise and a pair of counter-rotating vortices (hereafter termed linear profile), linear-log-law profile superposed with white noise and a pair of counter-rotating vortices (hereafter termed log profile), and synthetically generated three-dimensional flow field using the synthetic inflow generation method originally proposed by \cite{KimCastroXie2013} (hereafter termed synthetic profile). 
For clarity, a detailed discussion on the synthetic turbulence field generator has been included in Appendix A.
The linear profile initial condition based on the work by \cite{NelsonFringer2017} is given by

\begin{equation} \label{eq:linearprofile}
    u_i = 2 U_o \left( 1.0 - \frac{x_3}{H} \right) \delta_{i1} + \mathcal{U}(-\alpha U_b,\alpha U_b), 
\end{equation}

\noindent where $\mathcal{U}$ is the discrete analogue of the continuous uniform distribution, $\alpha$ is a tuning parameter set to a value of $0.7$ as recommended by \cite{NelsonFringer2017} for all cases. The scaled bulk velocity for the linear profile is given by \citep{NelsonFringer2017}

\begin{equation}
    U_o = \frac{u_{\tau}}{\kappa} \left[ \log \left(\frac{H}{z_o}\right) + \frac{z_o}{H} - 1 \right],
\end{equation}

\noindent where $z_o = \nu/(9u_{\tau})$. Two key differences must be noted between the \cite{NelsonFringer2017} formulation and the one used in this study: a. The linear profile is inverted in our case, b. No momentum forcing is used to keep the bulk mean velocity constant as done in \cite{NelsonFringer2017}. An inverse linear profile is used to effectively trigger the transition to turbulence as the shear stress at the bottom wall is large during the first eddy turnover.

The linear-log-law velocity profile is only applied to the streamwise velocity component given by

\begin{equation} \label{eq:linearloglaw}
    \begin{split}
    u_1 & = x_3  + \mathcal{U}(-\alpha U_b,\alpha U_b) \hspace{3cm} \forall \hspace{1mm} \frac{x_3 u_{\tau}}{\nu} \leq 11.6, \\
        & = \left( \frac{u_{\tau}}{\kappa} \log \left( \frac{x_3 u_{\tau}}{\nu} \right)+5.5 \right) + \mathcal{U}(-\alpha U_b,\alpha U_b) \hspace{5.5mm} \forall \hspace{1mm} \frac{x_3 u_{\tau}}{\nu} > 11.6,
    \end{split}
\end{equation}

\noindent while the pair of counter-rotating vortices are initialised by prescribing the spanwise and vertical velocity components following \cite{HenningsonKim1991}. The bulk dimensional velocity $U_b$ based on \cite{Pope2000} with a scaling factor of $0.5$ to avoid overshoot is given by

\begin{equation}
    U_b = 0.5 \left[ \frac{\nu}{H} \right] \left[ \frac{Re_{\tau}}{0.09} \right]^{\frac{1}{0.88}}.
\end{equation}

\noindent The choice for the two methods detailed in equations \ref{eq:linearprofile} and \ref{eq:linearloglaw} are motivated by previous work by \cite{NelsonFringer2017} and \cite{Costa2018}, which are the quickest ways to achieve a transition to turbulence followed by statistically stationary flow conditions.

\begin{figure}
    \centering
    \includegraphics[width=0.9\linewidth,trim={0cm 2.5cm 0cm 2cm},clip]{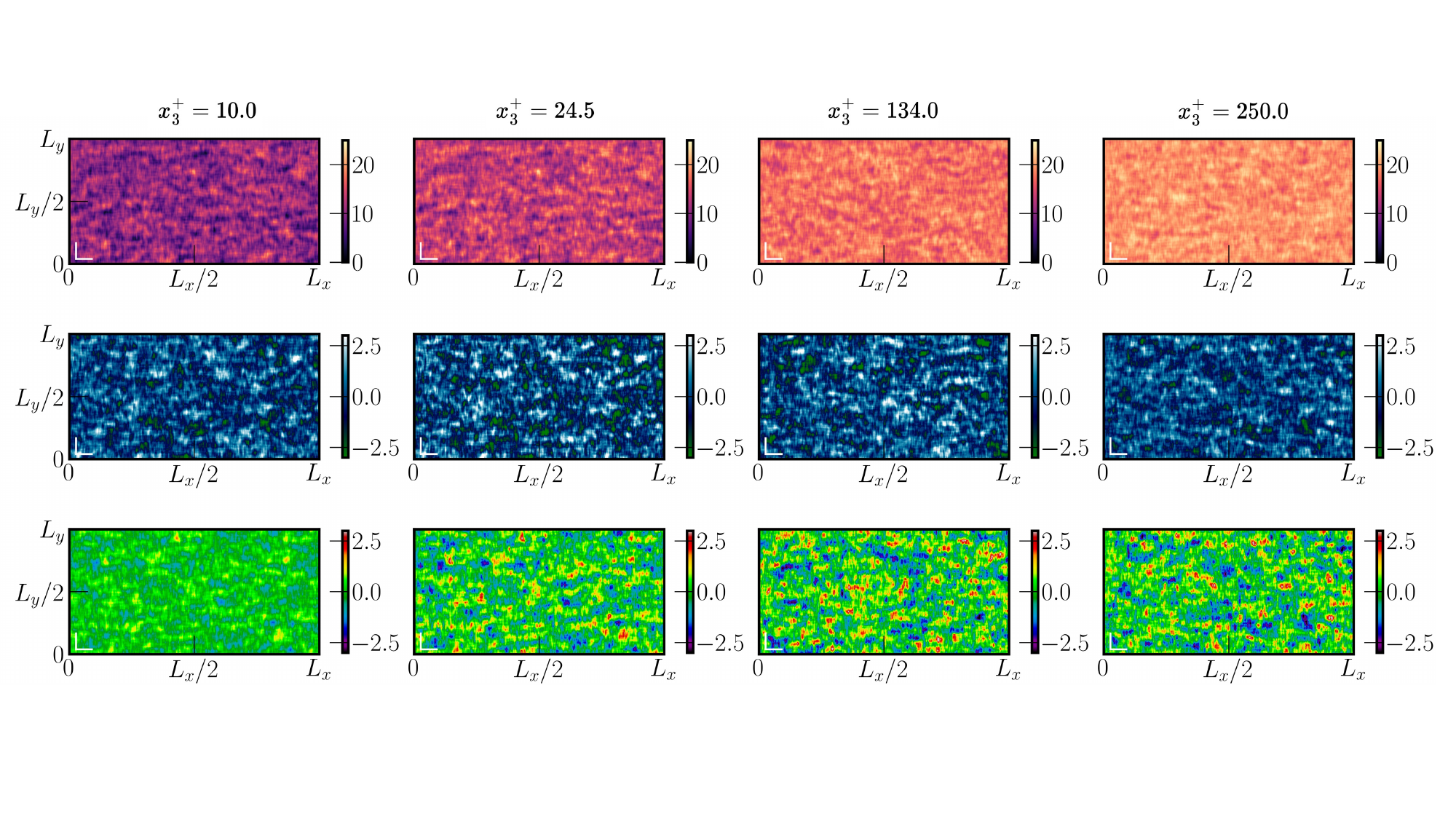}
    \caption{Snapshot of the initial condition generated using the STFG at various $x_3^+$ locations above the wall for $Re_{\tau} = 500$. The rows from top to bottom correspond to the streamwise, spanwise, and vertical velocities, respectively. Two white lines at the bottom left corner of each panel provide a reference length scale of $400$ wall units in the vertical and horizontal directions.}
    \label{fig:figure1}
\end{figure}

\begin{figure}
    \centering
    \includegraphics[width=0.8\linewidth,trim={0cm 0.5cm 0cm 0.5cm},clip]{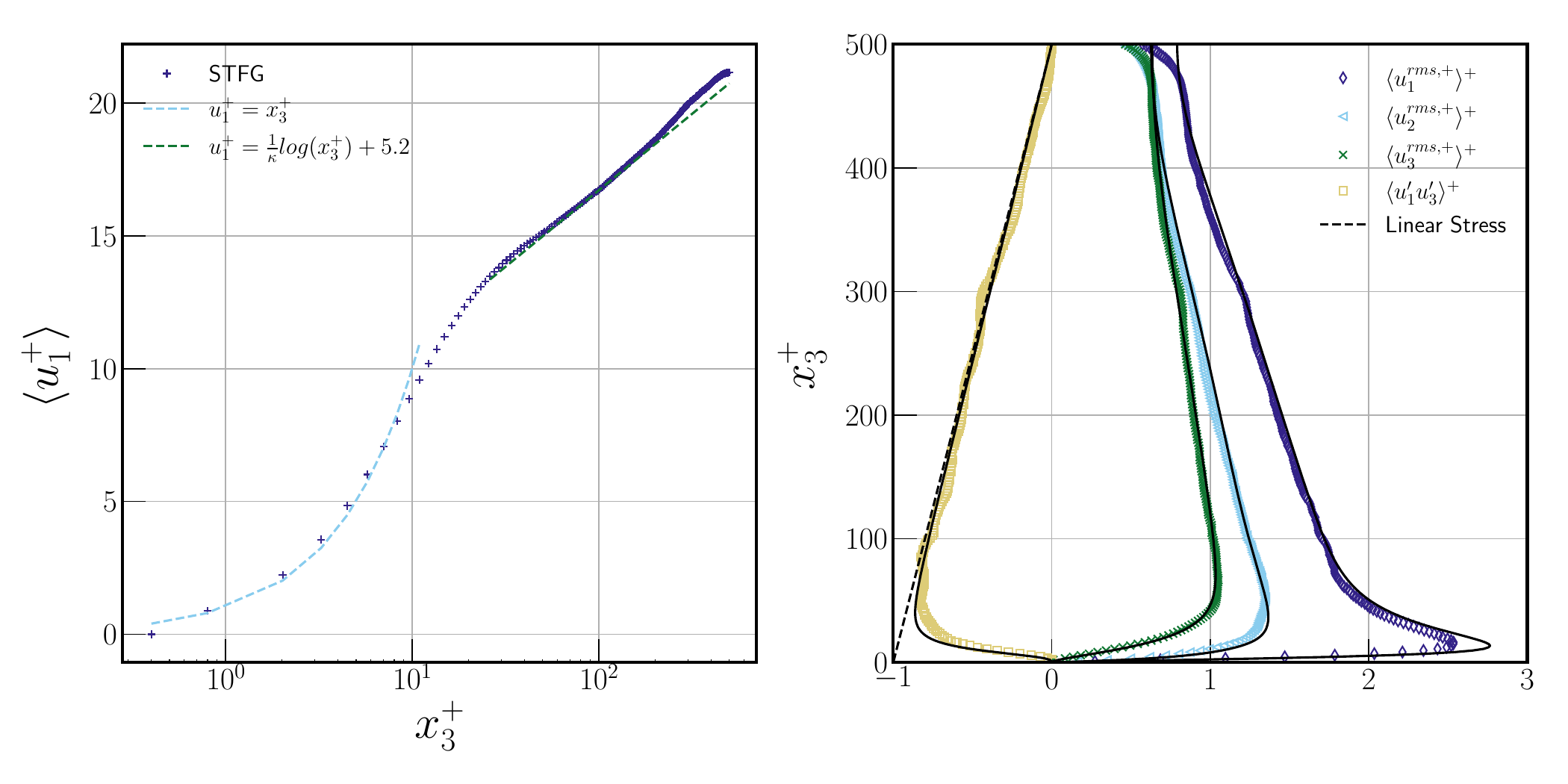}
    \caption{Planform averaged profile comparison of the STFG data against analytical solutions and reference data from \cite{MoserKimMansour1999} rescaled from $Re_{\tau} = 590$ to $Re_{\tau} = 500.0$.}
    \label{fig:figure2}
\end{figure}

The three-dimensional synthetic turbulent flow field is generated using the divergence-free method first proposed by \cite{KimCastroXie2013} as discussed in Appendix A.
For channel flows, the streamwise homogeneous direction allows one to exchange the spatial coordinate ($x_1$) and time ($t$) provided the right convective velocity is applied based on the frozen turbulence hypothesis \citep{Taylor1938}. 
This work uses the vertically integrated mean input velocity as the convective velocity ($U_c$). 
While the divergence-free form of the synthetic flow field is preferred, the first pressure-Poisson solution is sufficient to obtain such a divergence-free vector field; consequently, in this version of the synthetic turbulence field generator (STFG), only $U_c$ is matched at every time step during the signal sampling phase. 
Enforcing the divergence-free condition is relayed to the flow solver during the first time step without significantly increasing the computational cost.
The time step in the STFG is chosen based on the convection velocity and the grid spacing (known a-priori) given by

\begin{equation}
    \Delta t_{\text{STFG}} = \frac{\text{CFL} \Delta x_1}{U_c},
\end{equation}

\noindent where $\text{CFL}$ is the Courant-Friedrichs-Lewy number set to a value of $0.95$ unless specified otherwise both in the STFG and the simulation and $\Delta x_1$, is the grid spacing in the streamwise direction.
Figure \ref{fig:figure1} depicts the initial condition generated using the STFG, while Figure \ref{fig:figure2} shows the planform averaged profiles for the mean velocity and the root-mean-squared (rms) and Reynolds stress components compared against the data from \cite{MoserKimMansour1999}. 
Except for some minor differences observed in the streamwise rms velocity component, the STFG data corresponds well with the reference data.

To simplify the digital filter implementation, constant filter width is used in the spanwise and vertical directions despite the non-constant grid spacing used in the vertical direction.
In the current implementation of the synthetic flow generator, three user parameters are required: a. The target mean velocity profile $\overline{U}_1(x_3)$, b. The Reynolds stress tensor $R_{ij} (x_3)$, and c. Integral length scale $I_l$. 
A single scalar value for $I_l^+ \equiv I_l u_{\tau}/\nu = 100.0$ is applied isotropically for all the velocity components, unlike the method used in \cite{KimCastroXie2013}; this is primarily motivated by the need to keep the user input parameters as simple as possible. A serial implementation of the synthetic turbulent flow generator can be accessed through the open-source, public repository \href{https://github.com/AkshayPatil1994/SEM_KimCastroXie2013}{(Link to the public repository)}.

Table \ref{table:table1} details the various simulations carried out in this paper. 
All the simulations were run on the DelftBlue (DB) high-performance computing centre at the Delft University of Technology. 
Individual compute node consists of a 2 $\times$ 32-core  - Intel Xeon E5-6448Y 32C 2.1GHz processor, totalling 64 cores per node. 
Each of the $Re_{\tau} = 350$ cases required a total of 3392 CPU-hours to simulate 10 $T_{\epsilon}$ using 1 full node while each of the $Re_{\tau} = 500$ cases required a total of 13100 CPU-hours to simulate 10 $T_{\epsilon}$ using 2 full nodes. 
All simulations detailed in Table \ref{table:table1} were run for a total of 10$T_{\epsilon}$ and the temporal evolution of the statistics was compared by averaging the statistics for the last $5T_{\epsilon}$ unless mentioned otherwise.

\begin{table}[h!]
    \centering
    \begin{tabular}{c | c | c | c}
         Case Name & $Re_{\tau}$ & Initial Condition & Grid  \\
         \hline
         \hline
         Re350Lin & $350$ & Linear Profile & $1048 \times 764 \times 128 $ \\
         Re350Log & $350$ & Log Profile & $1048 \times 764 \times 128 $ \\
         Re350Syn & $350$ & Synthetic Profile & $1048 \times 764 \times 128 $ \\
         \hline
         Re500Lin & $500$ & Linear Profile  & $1500 \times 1048 \times 256 $ \\
         Re500Log & $500$ & Log Profile  & $1500 \times 1048 \times 256$ \\
         Re500Syn & $500$ & Synthetic Profile  & $1500 \times 1048 \times 256$  \\
         \hline
    \end{tabular}
    \caption{Simulation parameters considered in this paper. All simulations have a spatial resolution $\Delta x_1^+ = 4.18$, $\Delta x_2^+ = 3.0$, $\Delta x^+_{3,\text{min}} = 0.4$. For cases with $Re_{\tau} = 350$, $\Delta x^+_{3,\text{max}} = 5.4$ while for cases with $Re_{\tau} = 500$, $\Delta x^+_{3,\text{max}} = 3.7$. This resolution is sufficient for channel flows to resolve the requisite flow features of interest \citep{LozanoDuranJimenez2014}.}
    \label{table:table1}
\end{table}

\section{Results and Discussions}

\subsection{Convergence history for mean velocity and variances}

\begin{figure}[h!]
    \centering
    \includegraphics[width=0.8\linewidth,trim={0cm 0cm 0cm 0cm},clip]{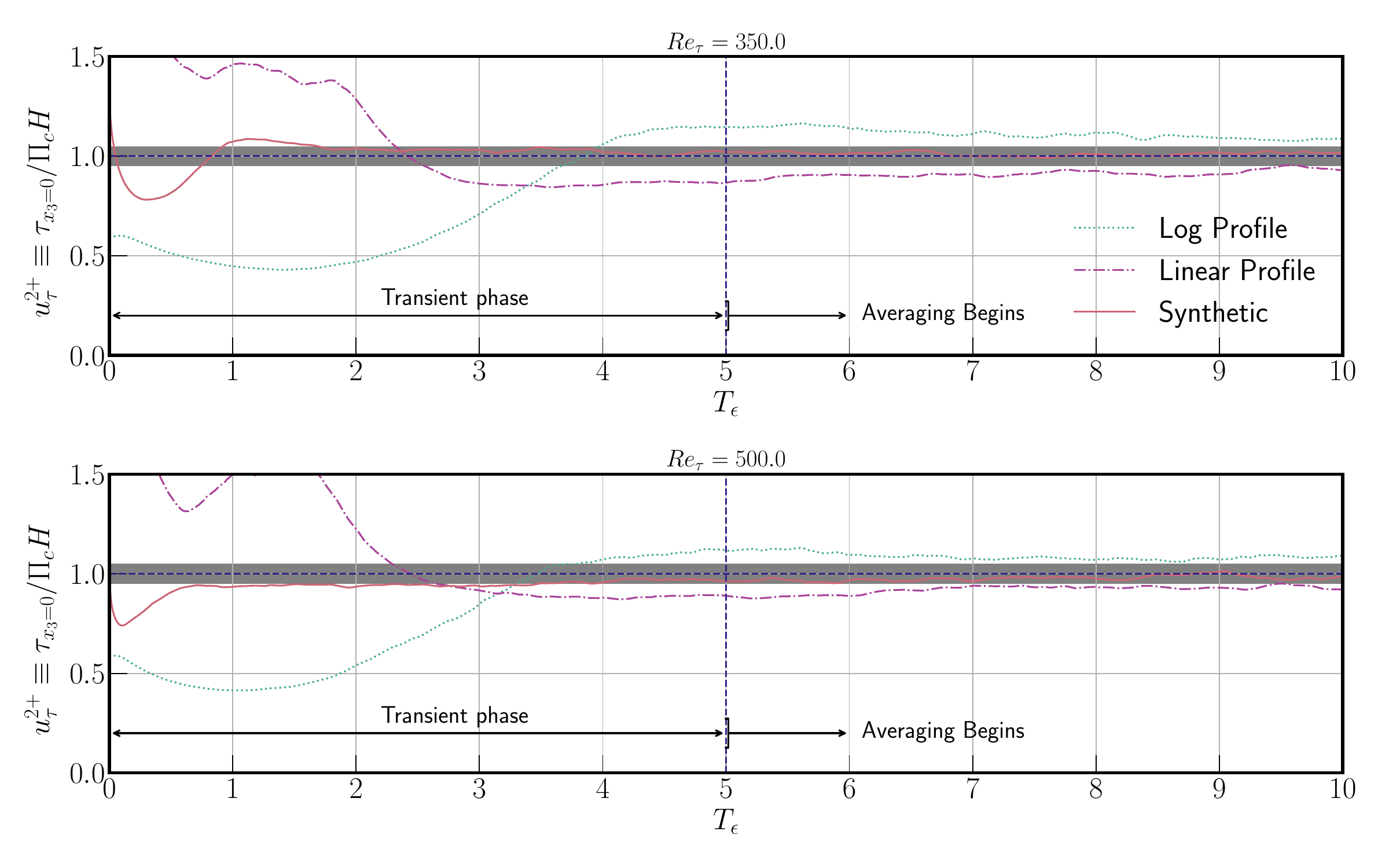}
    \caption{Comparison of the convergence of $u_{\tau}^2$ for the various initial conditions and $Re_{\tau}$ discussed in this study. The grey region marks the $\pm 5\%$ around the target value.}
    \label{fig:figure3}
\end{figure}

First, we present the convergence of the shear stress as a function of time, as it represents the global balance between the imposed pressure gradient and the balancing force therein. For planar channel flows driven by a constant pressure gradient, the shear stress at the wall balances the driving pressure gradient; thus, the shear/friction velocity ($u_{\tau}$) can be deduced from the driving pressure gradient and the channel height given by

\begin{equation} \label{eq:frictionVelocity}
    u_{\tau}^2 = \tau|_{x_3=0} = \nu \partial_3 \langle U_1 \rangle |_{x_3=0} = \Pi_c H,
\end{equation}

\noindent here it is assumed that the shear-stress has units of $m^2/s^2$ which is equivalent to setting $\tau = \tau/\rho_0$, where $\rho_0$ is the density of the fluid. Figure \ref{fig:figure3} compares the time evolution of the normalised shear stress for the three initial conditions and two $Re_{\tau}$ considered in this study. For cases with the log profile, the transition occurs identically for both values of $Re_{\tau}$ such that initially, the shear stress is approximately half of the target value, followed by a sudden transition to elevated shear stress due to the downward convecting pair of vortices that trigger the flow to a turbulent state. Despite scaling the vortex pair and the initial condition by the bulk velocity associated with the target flow Reynolds number, the shear stress still experiences an overshoot that requires more than $5 T_{\epsilon}$ to reach close to the $\pm 5\%$ of the target value. Comparing the behaviour of the vortex pair with the linear profile results in the opposite trend, where the shear stress has a large magnitude that effectively triggers the flow to transition to a turbulent state. However, once the transition occurs, the shear stress is lower than observed for the log profile cases. This difference can mainly be attributed to the fact that in the linear profile, once the flow transitions to turbulence, the largest velocities close to the wall are reduced, thus resulting in a smaller shear stress. In the log profile, the region close to the bottom wall exhibits a relatively higher velocity, thus resulting in an overall larger shear stress. As for the synthetically generated initial conditions, similar lower shear stress is observed within the first eddy turnover with a small overshoot above the target value for case Re350Syn, which eventually asymptotes around the target value. Since the synthetically generated turbulence retains the scaled components of all the fluctuating quantities, the flow is much closer to the target state and transitions to $\pm 5\%$ of the target value in $2T_{\epsilon}$. For case Re500Syn, it is observed that the shear stress is consistently smaller than the target value in the transient phase while being within the $\pm 5\%$ of the target value. However, the overall trend remains identical to case Re350Syn, despite this minor difference. Using a linear fit to the data after an initial transient of $5T_{\epsilon}$, the log profile is expected to converge at around $21.5 T_{\epsilon}$ and $19.0 T_{\epsilon}$, while the linear profile is expected to converge at $19.0 T_{\epsilon}$ and $15.0 T_{\epsilon}$ for the two $Re_{\tau}$'s presented in Figure \ref{fig:figure3}, respectively. The same linear fit suggests that for the log profile to first enter the $\pm5\%$ range, it would take a total of $\sim 11T_{\epsilon}$, while for the linear profile, it would take $\sim 12T_{\epsilon}$. A visual comparison can be seen in the supplementary materials, which compares the streamwise velocity along the centre of the domain in conjunction with the platform-averaged streamwise velocity profile. This discussion clearly illustrates the swift convergence to statistically stationary flow conditions for the synthetic initial condition compared to the other two considered in this study.

\begin{figure}[h!]
    \centering
    \includegraphics[width=0.9\linewidth,trim={4.5cm 0.2cm 2.5cm 0.2cm},clip]{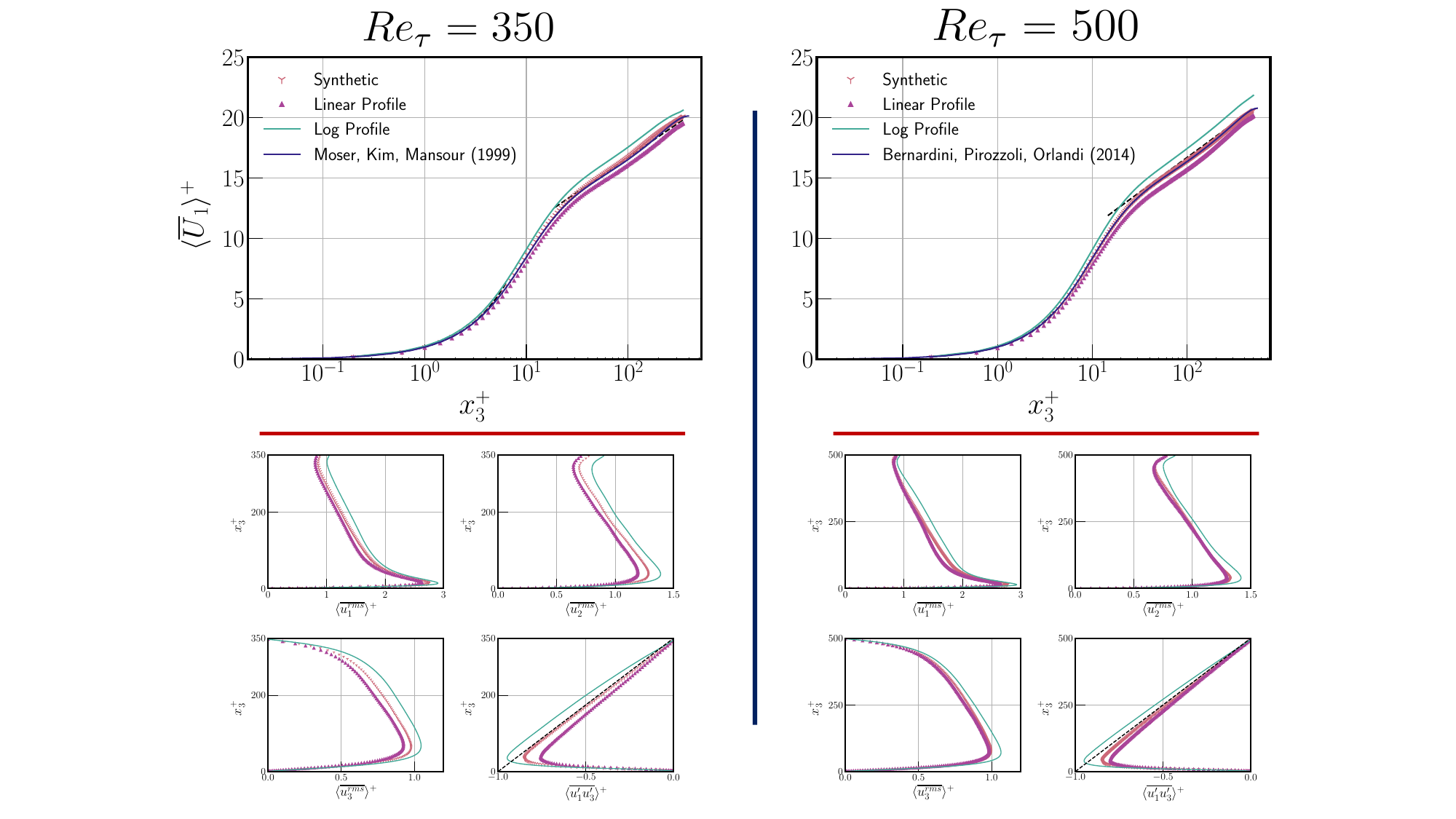}
    \caption{Comparison of the planform- and time-averaged mean velocity profiles and the planform- and time-averaged rms velocity and Reynolds stress profiles for the three initial conditions detailed in Table \ref{table:table1}. For $Re_{\tau} = 350$ velocity profile is compared against the \cite{MoserKimMansour1999} dataset with $Re_{\tau}=395$, while for $Re_{\tau} = 500$, the velocity profile is compared against the \cite{Bernardinietal2014} dataset with $Re_{\tau} = 550$.}
    \label{fig:figure4}
\end{figure}

Figure \ref{fig:figure4} compares the mean and rms velocity profiles for the three initial conditions detailed in the previous section. All profiles discussed below are averaged over the last $5T_{\epsilon}$ as detailed in Figure \ref{fig:figure3} with the right-pointing arrow. Since the mean velocity undergoes significant changes as a function of time during the transient phase, the time-averaged velocity changes as a function of the averaging window. Consequently, for a consistent comparison, the velocity is decomposed as

\begin{equation} \label{eq:velocityDecomposition}
    U_i (x_i,t) = \langle \overline{U}_i \rangle (x_3) + u_i^{\prime} (x_i,t),
\end{equation}

\noindent where, $\langle \overline{U}_i \rangle$ is the planform- and time-average using the last $5T_{\epsilon}$. While the convergence of the shear stress provides a first indication of the stationary state of the flow, the velocity profiles and their variances also need to converge to carry out meaningful averages. As seen in the planform- and time-averaged velocity ($\langle \overline{U}_1 \rangle$), excellent agreement between the expected analytical log-law (black dashed line) and case Re350Syn is observed as expected, while case Re350Log shows elevated velocity profile, and case Re350Lin shows slightly lower velocity profile. Case Re350Syn exhibits a slightly higher velocity magnitude in the inertial range which can be primarily attributed both to the relatively low Reynolds number and averaging time used to compute the statistics when compared to standard datasets. Regardless, swift statistically stationary conditions are expected based on the convergence history of the shear stress detailed in Figure \ref{fig:figure3}. The synthetic profile compares well with the data from \cite{MoserKimMansour1999}, further validating the convergence history observed in the shear stress. As case Re350Log shows elevated shear stress over the entire averaging period, the mean velocity profile suffers from an excess of total kinetic energy available that requires a longer time to be dissipated. For case Re350Lin, the opposite trend is observed and thus exhibits a relatively lower velocity magnitude. The rms velocity profiles and the Reynolds stress exhibit a consistent trend as observed in the mean velocity profile for all the cases with $Re_{\tau} = 350$.

For cases with $Re_{\tau} = 500$, a similar trend is observed when the three initial conditions are compared against each other. Specifically, case Re500Syn shows an exact match with the dataset from \cite{Bernardinietal2014} for the mean velocity profile, while cases Re500Log and Re500Lin exhibit a larger and smaller velocity magnitude when compared to the expected profiles, respectively. This trend between the three initial conditions is consistent across the rms velocity profiles and the Reynolds stress profiles, suggesting that the synthetic initial condition converges identically for higher friction Reynolds numbers. These observations collectively suggest that the mean velocity profiles and the variances converge relatively quickly for the synthetic initial condition compared to the analytical profiles discussed in this paper. In the following section, the convergence will be assessed for the higher-order statistics, such as the spectral energy and the turbulent kinetic energy budgets.

\subsection{Time evolution of the integral length scale and energy spectrum}

\begin{figure}[h!]
    \centering
    \includegraphics[width=0.8\linewidth,trim={0cm 0.5cm 0cm 0.5cm},clip]{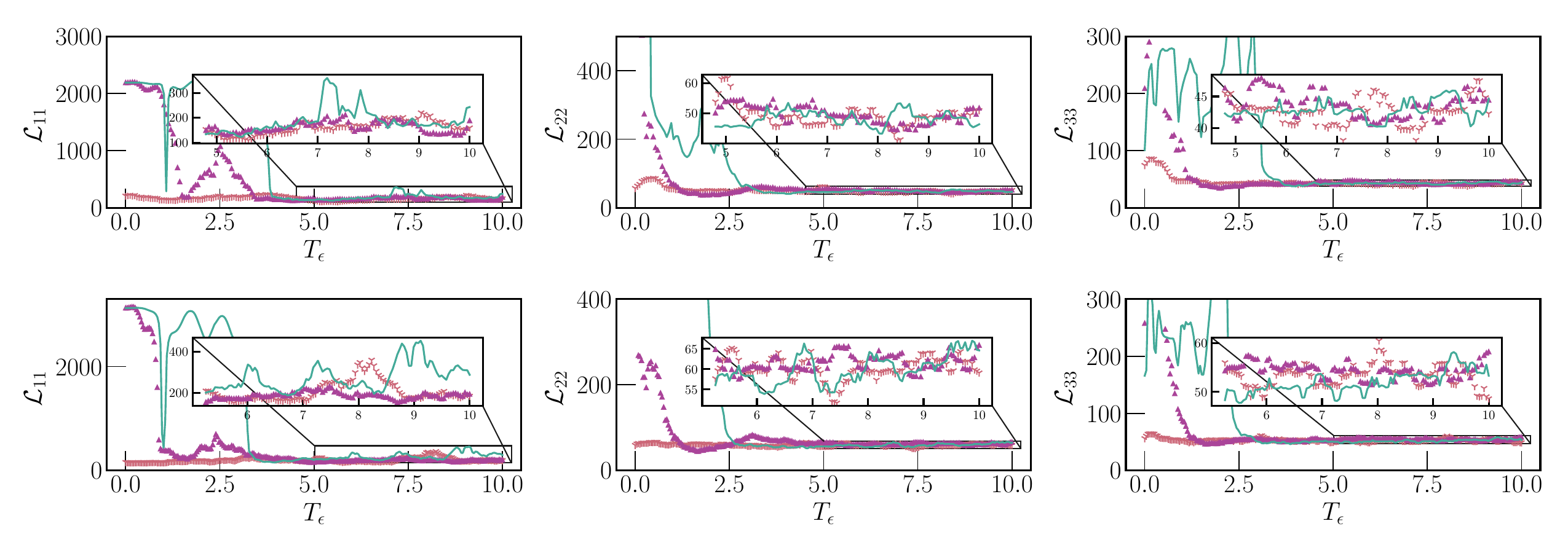}
    \caption{Comparison of spanwise-averaged integral length scale calculated using equation \ref{eq:integralLengthScale} for the two Reynolds numbers and three velocity components. The top row corresponds to $Re_{\tau} = 350$ at $x_3^+ = 204$ while the bottom row corresponds to $Re_{\tau} = 500$ at $x_3^+ = 275$. Data symbols are identical to the ones used in Figure \ref{fig:figure4}.}
    \label{fig:figure5}
\end{figure}

Auto-correlations along the streamwise direction provide a suitable metric for understanding the important scales of interest within the channel flow. Leveraging the auto-correlation of the three velocity components, the integral length scale for each velocity component is given by \citep{Tritton2012}

\begin{equation} \label{eq:integralLengthScale}
    \mathcal{L}_{\beta \beta}(x_3,t) = \int_{0}^{\mathcal{R_{\beta \beta}} < \frac{1}{e}} \mathcal{R_{\beta \beta}}(x_1,x_3,t) dx_1,
\end{equation}

\noindent where the repeating Greek indices represent the component of the velocity, while $\mathcal{R_{\beta \beta}}(x_1,x_3,t)$ is the spanwise averaged auto-correlation. Figure \ref{fig:figure5} compares the time evolution of the integral length scales for the three initial conditions and velocity components. For both the values of $Re_{\tau}$ with the synthetic initial condition, the magnitude for $\mathcal{L}_{11}$ is observed to be relatively steady. For case Re500Syn, there is a momentary increase in $\mathcal{L}_{11}$ around $8T_{\epsilon}$; however, this time-local increase in $\mathcal{L}_{11}$ is not significant and is quickly recovered to the statistically stationary value. Cases Re350Lin and Re500Lin undergo a similar change as the flow transitions to a turbulent state due to the large constant mean shear present over the entire channel depth, as detailed in Appendix B. For case Re350Lin, a secondary peak in $\mathcal{L}_{11}$ is observed at $2.5T_{\epsilon}$. In contrast, this peak for case Re500Lin is not as large in magnitude despite having identical transition behaviour as a function of time. Cases Re350Log and Re500Log exhibit similar transition behaviour when compared to each other. Despite a delayed onset of the turbulent flow conditions in cases Re350Log and Re500Log, the convergence to a statistically stationary condition occurs at approximately the same time as the linear profile.

The time-evolution for $\mathcal{L}_{22}$ and $\mathcal{L}_{33}$ largely confirm a similar trend where both the analytical profiles, i.e., linear and log, are initially observed to show large values that are an order of magnitude higher compared to the stationary conditions. Cases Re350Syn and Re500Syn are observed to undergo a relatively small initial transient where both $\mathcal{L}_{22}$ and $\mathcal{L}_{33}$ slightly increase compared to the stationary conditions; however, this transient is observed to quickly revert to the right order of magnitude within the first $T_{\epsilon}$, thereby oscillating around the stationary state. After the first $5T_{\epsilon}$, all the cases are observed to arrive at a similar value of integral length scales for all the components.

\begin{figure}[h!]
    \centering
    \includegraphics[width=0.8\linewidth,trim={0cm 2.0cm 0cm 1.5cm},clip]{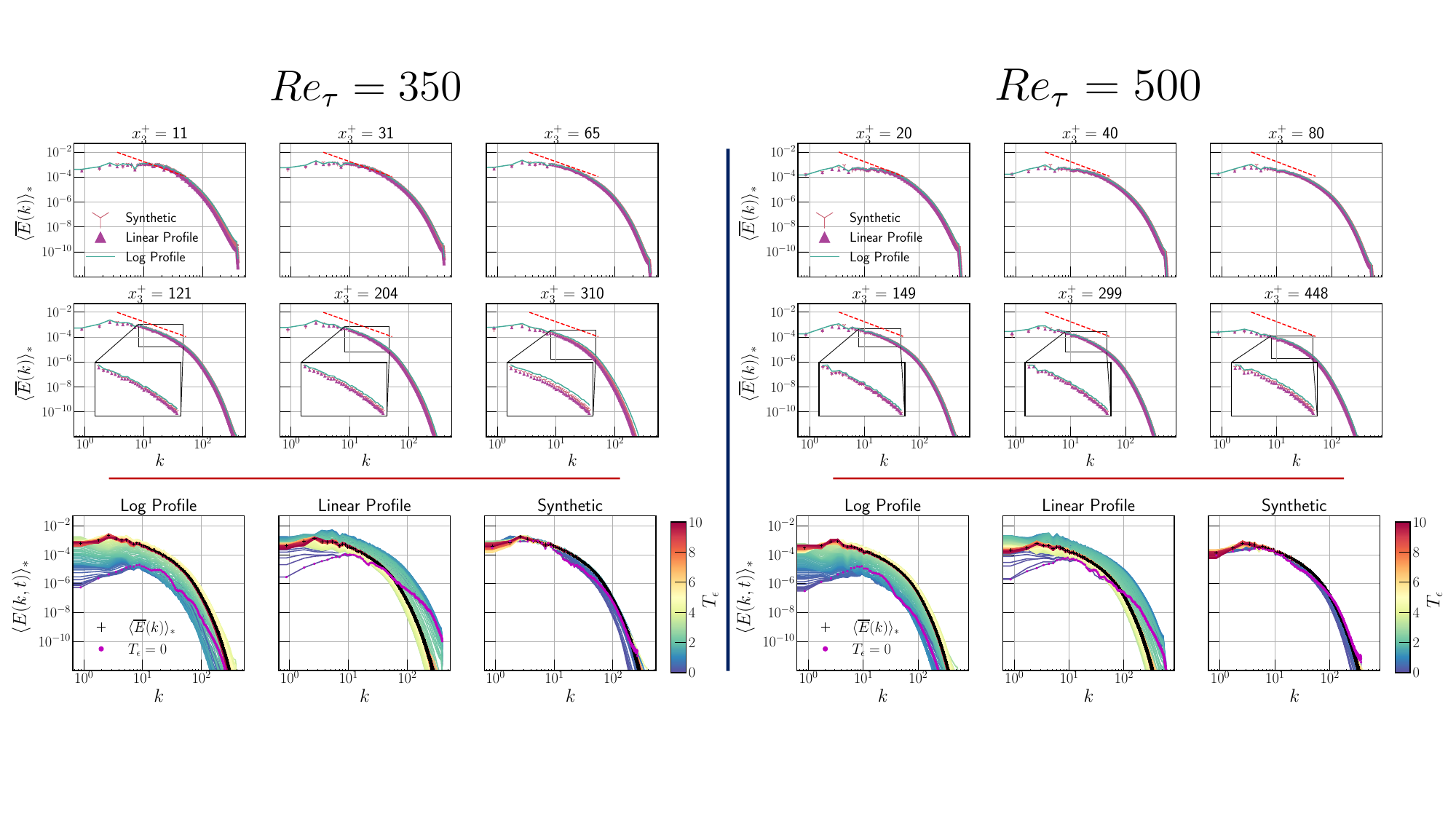}
    \caption{Comparison of spanwise- and time-averaged energy spectra for the cases discussed in this work. The top two rows compare the spectra at various wall-normal locations. The dashed-red line in the top two rows corresponds to the $-5/3^{\text{rd}}$ law. In contrast, the bottom rows, separated by the solid red line, compare the time evolution of the spectra at $x_3^+ = 204$ and $x_3^+ = 445$ for the three initial conditions discussed in this paper with $Re_{\tau} = 350$ and $Re_{\tau} = 500$, respectively. The $\langle \cdot \rangle_*$ operator signifies a planform-averaged quantity after the first $5T_{\epsilon}$ as marked in Figure \ref{fig:figure3}.}
    \label{fig:figure6}
\end{figure}

The integral length scale time evolution can be supported further by observing the time evolution of the energy spectrum. Specifically, in this case, we look at the planform-averaged spectrum for the velocity magnitude as detailed in Figure \ref{fig:figure6}. For the various wall-normal locations detailed in Figure \ref{fig:figure6}, a consistent trend between the three initial conditions is observed for both values of $Re_{\tau}$, where the linear profiles exhibit relatively smaller total energy at a given wavenumber compared to the synthetic and log-profiles, however, the difference between the three cases is not large. The time-evolution of the spectral energy for cases Re350Log and Re500Log has a consistent behaviour where the total energy during the first $5T_{\epsilon}$ is relatively lower, thereby increasing to a consistent value. This is also evidenced by supplementary movies 1 and 2 included along with this paper and the over time-evolution observed for the various parameters discussed earlier. The time evolution of the spectral energy for cases Re350Lin and Re500Lin also show a consistent transition to turbulence. A key difference between the linear and the log profiles is that the excursion around the mean value for the linear profile is relatively smaller compared to the one observed for the log profile case. As for cases Re350Syn and Re500Syn, they do not seem to deviate much from their initial state, and most of the changes are observed at the high wavenumber range. In contrast, the small wavenumbers corresponding to the large-scale turbulent features of the flow are relatively converged. Since the synthetic initial condition preserves the form of the Reynolds stress tensor using the exponential kernel in space, the distribution of spectral energy is preserved at the respective wavenumbers, unlike the analytical profiles where uncorrelated white noise is added on top of a mean profile. This is clearly seen in the spectrum shape at $T_{\epsilon} = 0$ compared to the averaged spectrum marked with black $+$ symbols.

\subsection{Turbulent Kinetic Energy Budget}

\begin{figure}[h!]
    \centering
    \includegraphics[width=0.8\linewidth,trim={0cm 0.5cm 0cm 0.5cm},clip]{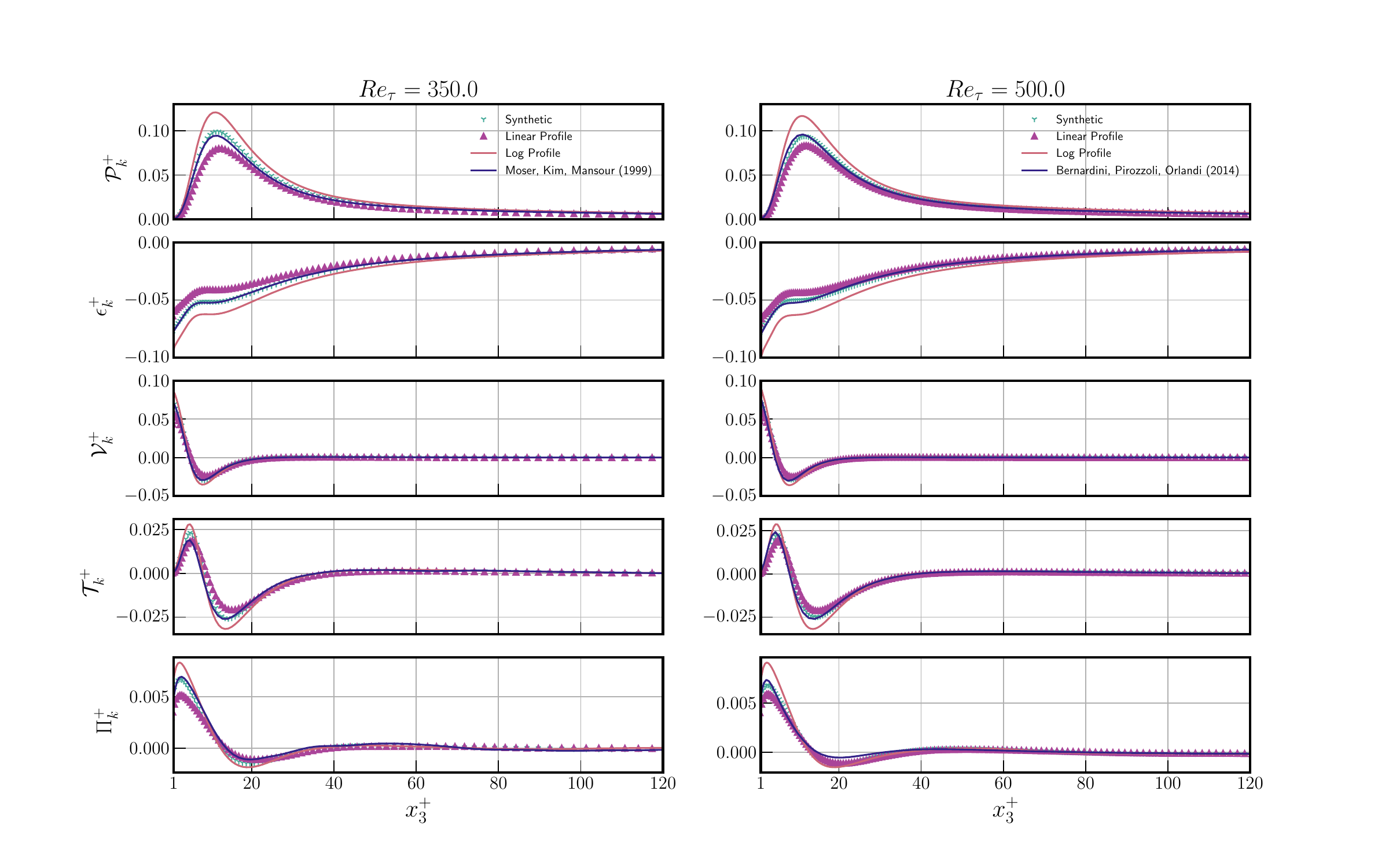}
    \caption{Comparison of the TKE budget terms where all the terms are non-dimensionalised using $u_{\tau}^4/( \kappa \nu)$.}
    \label{fig:figure7}
\end{figure}

The turbulent kinetic energy (TKE) budget for a channel flow with a streamwise driving pressure-gradient and homogeneous streamwise and spanwise directions is given by,

\begin{equation}
    \partial_t \mathcal{K} = \mathcal{P}_k - \mathcal{\epsilon}_k + \mathcal{V}_k - \mathcal{T}_k - \Pi_k,
\end{equation}

\noindent where $\mathcal{K} \equiv u_i^{\prime} u_i^{\prime}$ is the TKE, $\partial_t \mathcal{K}$ is the time rate of change of TKE and is zero for statistically stationary or steady-state conditions as will be assumed in this case, $\mathcal{P}_k$ is the production of TKE via mean shear, $\mathcal{\epsilon}_k$ is the dissipation rate of TKE, $\mathcal{V}_k$ is the viscous diffusion of TKE, $\mathcal{T}_k$ is the turbulent transport of TKE, and $\Pi_k$ is the pressure-diffusion of TKE, respectively. The last three terms are divergence terms and only transport the TKE within the domain without changing the net flux of TKE within the system. Figure \ref{fig:figure7} compares the various terms in the TKE budget that are averaged in the homogeneous directions and over $5T_{\epsilon}$ as indicated by the right arrow in Figure \ref{fig:figure3}. A consistent trend observed for the mean velocity and its variances is seen for the TKE budget terms; specifically, the TKE production, which is governed by the product of the Reynolds stress and the mean velocity gradient, has a larger magnitude for the log profile cases and a smaller magnitude for the linear cases. The synthetic initial condition is observed to follow the baseline reference data quite accurately, with case Re350Syn exhibiting a relatively small overshoot mainly because of the relatively larger mean velocity estimates compared to the reference dataset of \cite{MoserKimMansour1999}. The TKE dissipation rate agrees with the baselines dataset for both the Reynolds numbers quite accurately for the synthetic initial conditions. At the same time, the other two cases seem to differ quite substantially. The divergence terms in the TKE budget also compare well with the synthetic case despite the relatively short averaging and simulation window, suggesting that all the parameters of interest in the channel flow are adequately converged. 

\section{Conclusions and Recommendations}

In this study, we evaluated three different methods to initialise flow in a pressure-driven channel with the goal of achieving statistically stationary flow conditions in an efficient manner. Our results indicate that the synthetically generated three-dimensional turbulence method is the most computationally efficient and effective approach when compared against two analytically initialised velocity fields for reducing simulation spin-up time. The synthetic turbulence method, at a one-time computational cost of less than 1 CPU hour, achieved statistically stationary flow conditions within 3 eddy turnover times, a significant improvement compared to the alternative methods, which required more than 10 eddy turnovers. This reduction in spin-up time translates to substantial computational savings, making the synthetic turbulence approach particularly valuable when precursor turbulent initial conditions are unavailable. Additionally, we utilised a relatively simple synthetic turbulence generator and anticipate that using a more sophisticated method such as the ensemble synthetic method \citep{Schauetal2022} could potentially further reduce the spin-up time to reach statistically stationary flow conditions. The synthetically generated turbulence method offers a robust and resource-efficient strategy for setting up initial conditions in wall-bounded pressure-driven channel flow simulations when the right convective velocity is chosen. By minimizing the time to reach statistically stationary conditions, this approach enhances the efficiency of turbulence simulations, facilitating more rapid and cost-effective exploration of complex flow phenomena for such flow configurations.

\paragraph{Acknowledgements}
The authors acknowledge the use of computational resources of the DelftBlue supercomputer, provided by Delft High Performance Computing Centre (https://www.tudelft.nl/dhpc). AP would like to acknowledge the use of the colour-blind friendly python library cblind \citep{WafflardFernandezClement2023} that was used to make all the Figures in this manuscript.

\paragraph{Carbon Footprint Statement}
This work made use of the DelftBlue supercomputer and had an estimated footprint of 534~kg CO$_2$-equivalent (at least if not higher) using the Green Algorithms (\href{http://calculator.green-algorithms.org/}{http://calculator.green-algorithms.org/}). The input data used to arrive at these estimates were: Runtime - 773 hours, Number of cores - 64, Model Xeon X5660 (closest to the original hardware), and Memory available: 252 GiB located in the Netherlands. This is equivalent to taking 0.94 flight(s) from New York, U.S.A, to San Francisco, U.S.A.

\paragraph{Funding Statement}
This research was carried out as a part of the EU-Project RefMAP. RefMAP has received funding from the Horizon Europe program under grant agreement No 101096698. The opinions expressed herein reflect the authors’ views only. Under no circumstances shall the Horizon Europe program be responsible for any use that may be made of the information contained herein.

\paragraph{Declaration of Interests}
The authors declare no conflict of interest.

\paragraph{Author Contributions}
A.P. and C.G.S developed the research plan, and designed numerical 
experiments. A.P. did the data analysis and developed the code. A.P. and C.G.S. wrote the initial draft of the manuscript.

\paragraph{Data Availability Statement}
The synthetic turbulence generator can be accessed via the public repository \href{https://github.com/AkshayPatil1994/SEM_KimCastroXie2013}{{https://github.com/AkshayPatil1994/SEM\_KimCastroXie2013}}

\paragraph{Ethical Standards}
The research meets all ethical guidelines, including
adherence to the legal requirements of the study country. While preparing this manuscript, the generative AI tool Grammarly was used to check grammar and sentence structures.

\appendix

\section*{Appendix A: Synthetic Turbulence Field Generator based on \cite{KimCastroXie2013}}
\label{section:appendixA}

The synthetic turbulence field generator (STFG) is based on the divergence-free synthetic inflow generator first proposed by \cite{KimCastroXie2013}. This appendix provides a concise summary of the STFG while highlighting some of the differences in the implementation used in this paper. Much of this section is based on the work by \cite{KimCastroXie2013} and is repeated for reproducibility of the STFG code developed in this paper.

The velocity field $u_i$ is given by

\begin{equation} \label{eq:velocityVector}
    u_i = \overline{U}_i + a_{ij} u_{*,j},
\end{equation}

\noindent where $\overline{U}_i$ is the mean velocity profile known a-priori, $a_{ij}$ is the amplitude tensor, and $u_{*,j}$ is the unscaled fluctuations with zero mean, no correlation, and having unit variance. 
The amplitude tensor is obtained using the Cholesky decomposition of the Reynolds stress tensor ($R_{ij}$) and has a form given by \cite{Lundetal1998}

\begin{equation}
  a_{ij} = 
\begin{pmatrix}
  \sqrt{R_{11}} & 0.0 & 0.0 \\
  \frac{R_{21}}{a_{11}} & \sqrt{R_{22}-a_{21}^2} & 0.0 \\
  \frac{R_{31}}{a_{11}} & \left(\frac{R_{32} -a_{21}a_{31}}{a_{22}} \right) & \sqrt{R_{33} - a_{31}^2 - a_{32}^2}
\end{pmatrix}
\end{equation}

To generate the spatially correlated signal, a scalar field ($\psi_m$) is generated using a digital filter method given by

\begin{equation} \label{eq:psim}
    \psi_m = \sum_{j=-N}^{N} b_j r_{m+j},
\end{equation}

\noindent where $b_j$ is the model constant, $m$ and $j$ are position indices, $N=2n$, $n=I_l/\Delta x_i$, $I_l$ is the integral length scale, and $\Delta x_i$ is the grid spacing in the coordinate directions. 
Unlike the work by \cite{KimCastroXie2013}, in this version of the STFG, the integral length scale is assumed to be isotropic in all directions for user input and implementation simplicity. 
Here, $\phi_m$ constitutes a one-dimensional array with zero mean, unit variance, and spatially correlated signal. 
The model constant $b_j$ is given by

\begin{equation}
    b_j = \frac{b_j^{\prime}}{\sqrt{\sum_{l=-N}^{N} b_l^{{\prime}2}}},
\end{equation}

\noindent and $b_j^{\prime} = \exp{\left( -\frac{\pi | j| }{2n} \right)}$. In most cases, the flow problem will be solved in three dimensions; thus, a two-dimensional extension of equation \ref{eq:psim} can be formulated as

\begin{equation} \label{eq:2dpsi}
    \psi_{m,l} = \sum_{j=-N}^{N} \sum_{k=-N}^{N} b_j b_k r_{m+j,l+k}.
\end{equation}

Using the spatially correlated two-dimensional data as detailed in equation \ref{eq:2dpsi}, the temporal correlations are implemented through the specification of $u_{*,i}$ given by

\begin{equation}
    u_{*,i}(t + \Delta t) = u_{*,i} (t) \exp{\left( -\frac{C_{XC}\Delta t}{T} \right)} + \psi_i (t) \left[ 1 - \exp{\left( -\frac{2C_{XC}\Delta t}{T} \right)} \right]^{\frac{1}{2}},
\end{equation}

\noindent where $T = I_l/U_c$ is the Lagrangian time scale, $U_c$ is the convective velocity defined in this case as the vertically integrated bulk velocity ($U_c = H^{-1} \int_{x'_3=0}^{x'_3=H} \overline{U}_i dx'_3$). 
Using a constant correction of the mass-flux for the synthetically generated velocity $u_i$, a divergence-free form of the velocity field can be supplied.

\section*{Appendix B: Mean shear profiles}
\label{section:appendixB}

\begin{figure}[h!]
    \centering
    \includegraphics[width=0.8\linewidth,trim={0cm 0.0cm 0cm 0.0cm},clip]{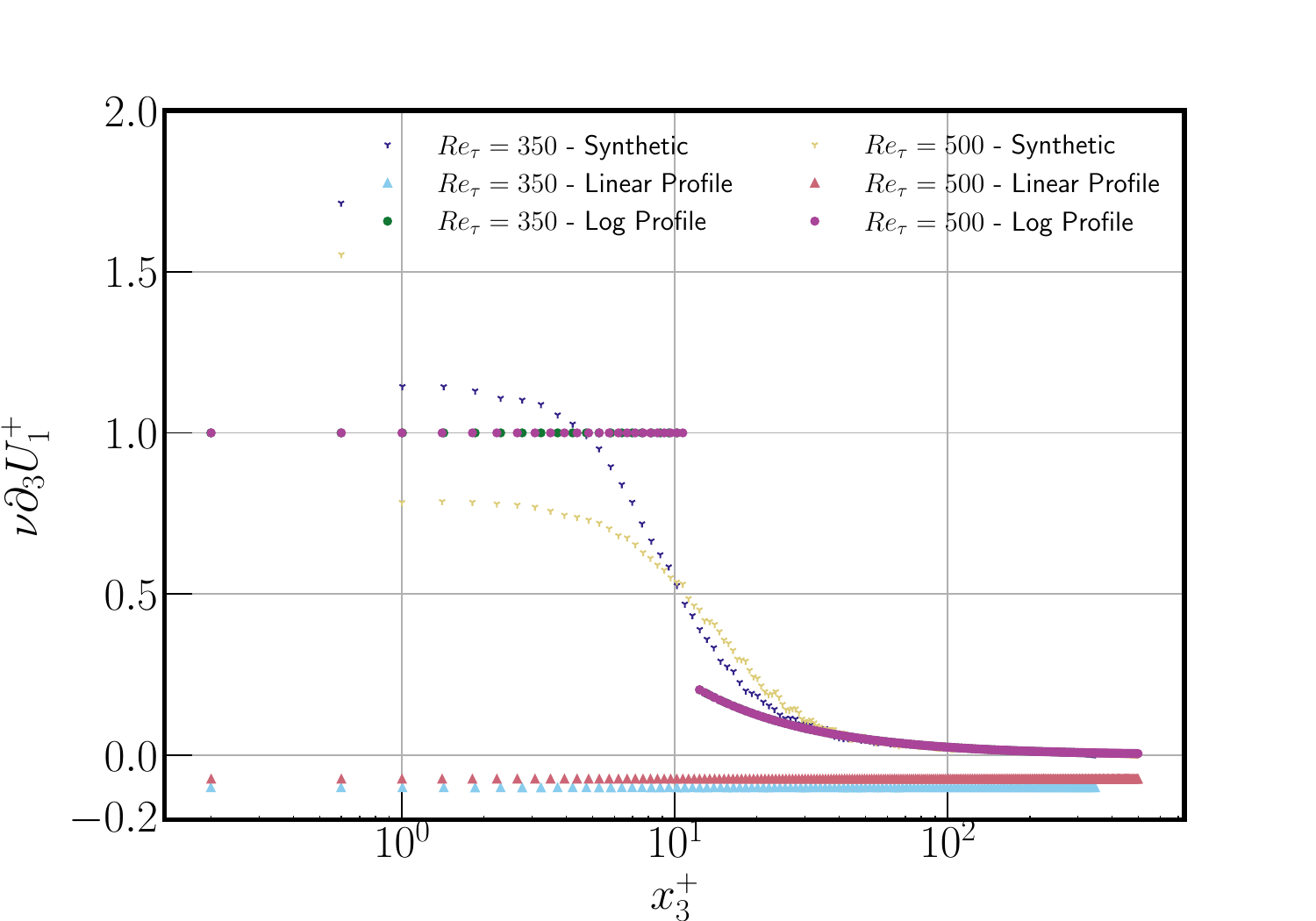}
    \caption{Comparison of the initial shear stress as a function of the wall-normal direction for the various cases simulated in this paper.}
    \label{fig:appendix_figure1}
\end{figure}

Based on the analytical expressions for the initial conditions, it is possible to a-priori estimate the mean shear profiles used to set up the initial conditions. Since the channel is homogeneous in the streamwise and spanwise directions, the mean shear is non-zero for the vertical gradient of the streamwise component (i.e., $\partial_3 U_1$). For the linear profile, the shear stress that is provided as the initial condition is given by

\begin{equation}
    \nu \partial_3 U_1 = \left[ \frac{-2 u_{\tau}^2}{\kappa} \right] \left[  \frac{1}{Re_{\tau}} \right] \left[ \log \left( \frac{H}{z_o} \right) + \frac{z_o}{H}  - 1\right],
\end{equation}

\noindent while for the linear-log-law profile, the shear stress is given by

\begin{equation}
    \begin{split}
    \nu \partial_3 U_1 & = \nu, \hspace{3.4cm} \forall \frac{x_3 u_{\tau}}{\nu} \leq 11.6 \\
                       & = \left[ \frac{u_{\tau}^2}{\kappa} \right] \left[ \frac{\nu}{u_{\tau} x_3} \right] = \left[ \frac{u_{\tau}^2}{\kappa} \right] \left[ \frac{1}{Re_{\tau}^{*}} \right], \hspace{2cm} \forall \frac{x_3 u_{\tau}}{\nu} > 11.6
    \end{split}
\end{equation}

\noindent where $Re_{\tau}^{*} \equiv u_{\tau} x_3/\nu$ is the local friction Reynolds number. Figure \ref{fig:appendix_figure1} compares the stress profiles for the three initial condition types and two friction Reynolds numbers. For the linear profile, a constant stress magnitude is applied, while for the linear-log-law profile, the stress varies as a function of the distance away from the wall.

\bibliography{paper.bib}

\end{document}